\documentclass[conference]{IEEEtran}
\usepackage{amsthm}
\usepackage{amsmath}
\usepackage{cases}
\usepackage{graphicx}
\usepackage{CJK}

\ifCLASSINFOpdf
\else
\fi
\begin{document}

\newtheorem{Thm}{\textbf{Theorem}}
\newtheorem{Lem}{\textbf{Lemma}}
\newtheorem{Def}{\textbf{Definition}}
\newtheorem{Rem}{\textbf{Remark}}
\newtheorem{Exam}{\textbf{Example}}
\newtheorem{Sup}{\textbf{Assumption}}
\newtheorem{Cor}{\textbf{Collary}}
\title{ Focusing on a Probability Element: Parameter Selection of Message Importance \\Measure in Big Data}


%
\author{\IEEEauthorblockN{Rui She\IEEEauthorrefmark{1},
Shanyun Liu\IEEEauthorrefmark{1},
Yunquan Dong\IEEEauthorrefmark{2},
and Pingyi Fan\IEEEauthorrefmark{1}, $Senior Member$, \textsl{IEEE}
}
\IEEEauthorblockA{\IEEEauthorrefmark{1}
State Key Laboratory on Microwave and Digital Communications\\
Tsinghua National Laboratory for Information Science and Technology\\
Department of Electronic Engineering, Tsinghua University, Beijing,
P.R. China\\
Emails: \{sher15, liushany16\}@mails.tsinghua.edu.cn, fpy@tsinghua.edu.cn
}
\IEEEauthorblockA{\IEEEauthorrefmark{2}School of Electronic and Information Engineering,\\
Nanjing University of Information Science and Technology, Nanjing, P.R. China\\
Email: yunquandong@nuist.edu.cn
}
}


\maketitle

\begin{abstract}
Message importance measure (MIM) is applicable to characterize the importance of information in the scenario of big data, similar to entropy in information theory. In fact, MIM with a variable parameter can make an effect on the characterization of distribution.
Furthermore, by choosing an appropriate parameter of MIM, it is possible to emphasize the message importance of a certain probability element in a distribution. Therefore, parametric MIM can play a vital role in anomaly detection of big data by focusing on probability of an anomalous event.
In this paper, we propose a parameter selection method of MIM focusing on a probability element and then present its major properties. In addition, we discuss the parameter selection with prior probability, and investigate the availability in a statistical processing model of big data for anomaly detection problem.
\end{abstract}


%
\IEEEpeerreviewmaketitle

\section{Introduction}
Recently, big data has attracted considerable attention in many fields of applications. In particular, there are some cases where the minority subsets are more important than those majority subsets. For instance, in Intrusion Detection Systems (IDSs), only a few signs of security violations, which give rise to alarms, should be monitored and analyzed \cite{Data Mining Approaches for Intrusion Detection,Mining intrusion detection alarms for actionable knowledge}. What is more, as far as financial crime detection is concerned, the identities not conforming to the provisions which can contribute to financial frauds, are certainly in the minority \cite{A comprehensive survey of data mining-based accounting-fraud detection research}. Hidden in big data sets \cite{Efficient algorithms for mining outliers from large data sets}, the minority subset detection or anomaly detection becomes increasingly influential in various fields. \par

Note that the conventional information theoretic measures such as Kolomogorov Complexity, Shannon entropy and relative entropy still applied to detect infrequent and exceptional appearance in large data sets.

By judging the complexity of classified big data sets using these measures, anomalous subset could be recognized \cite{Anomaly detection: A survey}, \cite{Information-theoretic measures for anomaly detection}.

Moreover, with an information theoretic approach, the objective function related to factorization based on distribution was constructed to detect minority subset \cite{An information theoretic approach to detection of minority subsets in database}. Although these algorithms are suitable for some special applications, it is not difficult to perceive that they actually focus on processing the majority subsets.\par

As a new information measure, message importance measure(MIM) has been proposed to target minority subsets in big data opposed to conventional information theoretic measures \cite{Elements of information theory 2nd edition}, \cite{Message Importance Measure and Its Application to Minority Subset Detection in Big Data}. In particular, this information measure is similar to Shannon entropy and Renyi entropy in structure, while focusing more on significance of the anomaly with small occurring probability in big data scenarios. In terms of applications of MIM, it was conducive to the minority subset detection including binary minority subset detection and M-ary minority subset detection \cite{Message Importance Measure and Its Application to Minority Subset Detection in Big Data}. However, the minimum occurring probability of anomalous events attracts the most attention to select the parameter of MIM in the previous method. That is to say, other anomalous events are ignored except the one with minimum probability. In addition, the previous parameter selection principle of MIM is too imprecise to set an operational parameter of message importance measure for a given distribution. Thus, it is necessary to present a practical parameter selection which can also focus on anomalous events with other small probability in addition to the one with the minimum probability.

Objectively, MIM can present an efficient implement to address the problem of anomalous events detection. To make MIM adaptable to detect anomalies in big data era, we will introduce the parameter selection method of MIM focusing on a probability that maintains availability to anomaly detection problem in big data analysis.\par

\subsection{Parameter Selection of MIM }
In this subsection, we shall propose a parameter selection of MIM to pay close attention to importance of a probability element in a distribution. Before we proceed, let us review the definition of MIM briefly first \cite{Message Importance Measure and Its Application to Minority Subset Detection in Big Data}.\par
In a finite alphabet, for a given probability distribution $\textbf{\emph{p}}=( p_1, p_2,..., p_n)$, the MIM with importance coefficient $\varpi \geq 0$ is defined as

\begin{equation}
L(\textbf{\emph{p} },\varpi) = \log\left( \sum_{i=1}^n p_ie^{\varpi\left(1-p_i\right)}\right),
\end{equation}
which measures the information importance of the distribution.

\begin{Def}
For a finite distribution $\textbf{\textsl{p}}=( p_1, p_2,..., p_n)$, the parameter selection of MIM focusing on a certain probability element, which is a functional relationship between the importance coefficient $\varpi$ and a given probability element $p_j$, is defined as

\begin{equation}
\begin{aligned}
\varpi_{j}= F\left (p_j \right )
\end{aligned}
\end{equation}
and the corresponding MIM is
\begin{equation}
\begin{aligned}
L_{j}\left (\textbf{p} , \varpi_{j} \right ) = \log\left( \sum_{i=1}^n p_ie^{\varpi_{j}\left(1-p_i\right)}\right),
\end{aligned}
\end{equation}
where importance coefficient $\varpi$ can be set by $p_j$.
\end{Def}

\subsection{ Outline of the Paper }
The rest of this paper is organized as follows. In Section II, we present a parameter selection method of MIM focusing on a certain probability element and analyze the properties of the message importance measure. In Section III, we perform a detailed discussion on the parameter selection of MIM. Then the availability of parametric MIM is discussed in Section IV by applying the parametric MIM in a model of minority subset, which is closely related to the applications in big data. We present some simulation results to verify our discussions in Section V. Finally, we conclude the paper in Section VI.

\section{ A MIM with probability focusing and its properties}
In this section, we shall propose a parameter selection method of MIM which can focus on the importance of a certain probability element. In order to gain an insight into the message importance measure with this method, its fundamental properties are investigated in depth.

Similar to the self-information in Shannon entropy, an importance coefficient selection method of MIM focusing on a given probability $p_j$ is defined as
\begin{equation}
\varpi_{j}=  F(p_j)=\frac{1}{ p_j }.
\end{equation}
In accordance with the functional relationship between importance coefficient $\varpi$ and probability element $p_j$, the parametric MIM is given by
\begin{equation}
\begin{aligned}
L_{j}\left (\textbf{\emph{p} },\varpi_{j} \right ) = & \log\left( \sum_{i=1}^n p_ie^{\frac{1}{p_j}\left(1-p_i\right)}\right). \\
\end{aligned}
\end{equation}
Moreover, some properties of the MIM can be discussed to provide some possibly theoretical guidance for applications as follows:

\textsl{1) Principal Component: }For a fixed distribution $\textbf{\emph{p}}=( p_1, p_2,..., p_n)$ with $p_j > 0 \ (j=1,...,n)$, when we focus on the importance of a certain probability element $p_j$, the probability $p_j$ becomes the principal component in the message importance measure. That is

\begin{equation}
p_{j}e^{\varpi_j (1-p_j)} \geq  p_{i}e^{\varpi_{j}(1-p_i)}     \ \ \left( i=1,...,n \right).
\end{equation}

\textsl{proof: }Consider the derivative
\begin{equation}
\begin{aligned}
\frac{\partial \left( pe^{\varpi (1-p)} \right)}{\partial p}
 = \left(1-\varpi p \right) e^{\varpi (1-p) },
\end{aligned}
\end{equation}
where $\varpi=\frac{ 1 }{ p_j }$. We note that when $p=p_j$, the derivative is zero and the quadratic derivative is negative. Thus the property is proved.

\textsl{2) Chain Rule for MIM: }For a given distribution $\textbf{\emph{p}}=( p_1, p_2,..., p_n)$ without $p_j=0$, if $p_1 > p_2 > ... > p_n$ , then we have
\begin{equation}
L_1(\textbf{\emph{p} },\varpi_{ 1 }) < L_2(\textbf{\emph{p} },\varpi_{ 2 }) < ... < L_n(\textbf{\emph{p} },\varpi_{ n }).
\end{equation}

\textsl{proof: }Since $p_1 > p_2 > ... > p_n$ and the importance coefficient is $\varpi_{j}=\frac{1}{p_j} \ \left(j=1,...,n\right)$, we have $\varpi_{ 1 } < \varpi_{ 2 } < ... <\varpi_{n} $.

This derivative of $L(\textbf{\emph{p} },\varpi)$ with respect to $\varpi$ is
\begin{equation}
\begin{aligned}
&\frac{\partial L(\textbf{\emph{p} },\varpi)}{\partial \varpi}\\
& = \frac{  \sum_{i=1}^n p_ie^{\varpi\left(1-p_i\right)}-\sum_{i=1}^n p_i^2e^{\varpi\left(1-p_i\right)} }{  \sum_{i=1}^n p_ie^{\varpi\left(1-p_i\right)} }\\
& =1- \frac{ \sum_{i=1}^n p_i^2e^{\varpi\left(1-p_i\right)} }{ \sum_{i=1}^n p_ie^{\varpi\left(1-p_i\right)} } >0,
\end{aligned}
\end{equation}
which means $L(\textbf{\emph{p} },\varpi)$ is monotonically increasing with $\varpi$ and the property is proved.

\textsl{3) The Lower Bound for Finite Distribution: }For any finite distribution $\textbf{\emph{p}}=( p_1, p_2,..., p_n)$ with positive $p_j$, we denote $p_{max}$ as the maximum $p_j$, then we have
\begin{equation}
\begin{aligned}
&L_j\left (\textbf{\emph{p} },\varpi_{j}\right) \\
&\geq L\left (\textbf{\emph{p} },\frac{1}{ p_{max} }\right )\\
&> L\left (\textbf{\emph{p} },1\right )=\log\left( \sum_{i=1}^n p_ie^{\left(-p_i\right)}\right)+1.
\end{aligned}
\end{equation}
This property follows from the chain rule property and the proof is omitted.

\textsl{4) Focusing on the Minimum Probability: }For any $\textbf{\emph{p}}=( p_1, p_2,..., p_n)$ without zeros element, denote $p_{min}$ as the minimum probability element and $\varpi_{0} = \frac{ 1 }{ p_{min}}$ according to the parameter selection method. In that case, we have

\begin{equation}
\begin{aligned}
L_{0}\left ( \textbf{\emph{p} },\varpi_{0} \right ) \geq L_0\left ( \textbf{\emph{u} },\varpi_{0} \right ),
\end{aligned}
\end{equation}
where $\textbf{\emph{u}}$ is the uniform distribution, with the same number of elements as  $\textbf{\emph{p}}$.

\textsl{proof: } For the binomial distribution $\textbf{\emph{p}}=( p, 1-p)$ with $ 0< p < \frac{1}{2}$, we have $\varpi_{0}=\frac{1}{p}$. In order to compare the two entire parametric MIM apparently, we define a function $G(\textbf{\emph{p}})=
L( \textbf{\emph{p} },\varpi_{0})-L( \textbf{\emph{u} },\varpi_{0})$. Then we can derive
\begin{equation}
\begin{aligned}
G\left (\emph{p}\right )= \log\left ( pe^{\frac{1}{p}-2}+(1-p) \right  ),
\end{aligned}
\end{equation}

\begin{equation}
\begin{aligned}
\frac{ \partial G(\emph{p})}{\partial p}=
\frac{ (1-\frac{1}{p})e^{\frac{1}{p}-2}-1 }{  pe^{\frac{1}{p}-2}+(1-p) }<0.
\end{aligned}
\end{equation}
It is clear that $G(\emph{p}) \geq G(\emph{ p} = \frac{1}{2} ) =0$, which certifies the property.\par

Next for the any distribution $\textbf{\emph{p}}=( p_1, p_2,..., p_n)$ without zero element, we have $\varpi_{0}=\frac{1}{p_{min}}$. We define the Lagrange function
\begin{equation}
\begin{aligned}
G\left (\textbf{\emph{p}},\lambda\right )&= G\left ({p_1},....,{p_n},\lambda \right )\\
&= \sum\limits_{i = 1}^n {{p_i}{e^{\frac{{1 - {p_i}}}{{{p_{\min }}}} - n + 1}}}  + \lambda \left (\sum\limits_{i = 1}^n {{p_i}}  - 1 \right ).
\end{aligned}
\end{equation}
Then the partial derivative of $G(\textbf{\emph{p}},\lambda)$ with regard to $p_i$ is obtained as follows
\begin{equation}
\begin{aligned}
\frac{{\partial G}}{{\partial {p_i}}} = \left (1 - \frac{{{p_i}}}{{{p_{\min }}}} \right ){e^{\frac{{1 - {p_i}}}{{{p_{\min }}}} - n + 1}} + \lambda.
\end{aligned}
\end{equation}
By solving $\frac{{\partial G}}{{\partial {p_i}}}=0$ with $\sum\limits_{i = 1}^n {{p_i}}  - 1 = 0$, it can be readily demonstrated that $p_1=p_2=...=p_n=\frac{1}{n}$ is the solution, which denotes the extreme value of \textsl{$G$} will be obtained by the uniform distribution.

In addition, we can attain

\begin{equation}
\begin{aligned}
&L_0\left ( \textbf{\emph{p} },\varpi_{0} \right )-L_0\left ( \textbf{\emph{u} },\varpi_{0} \right )\\
&=\sum\limits_{i = 1}^n {{p_i}{e^{\frac{{1 - {p_i}}}{{{p_{\min }}}} - n + 1}}}\\
&= \log \left ( G \left ({p_1},....,{p_n},\lambda=0 \right ) \right ) \\
\end{aligned}
\end{equation}

\begin{equation}
G({p_1},....,{p_n},\lambda=0 ) \geq 1.
\end{equation}
Therefore
$L_0\left ( \textbf{\emph{p} },\varpi_{0} \right )-L_0\left ( \textbf{\emph{u} },\varpi_{0} \right ) \geq 0$.

\section{ Parameter selection with prior probability }
In this section, we investigate the parameter selection of message importance measure when the range of prior probability can be predicted. In that regard, there are certainly many scenarios where the prior probability of anomalies is too small to be estimated. In order to reduce the complexity of the parameter selection approach of message importance measure when an interested probability is focused on, we will discuss binary distribution in the scenario with two possible events.

\subsection{Parameter Focusing on Prior Probability}
 For a given binomial distribution $\textbf{\emph{p}}=( p_1, p_2)=( p,1-p )$, with the prior probability $ p_{i}^{(1)}< p_i <p_{i}^{(2)}, (i=1, 2)$, $0< p < \frac{1}{2}$, we mean to select an appropriate importance coefficient $\varpi$ to achieve the entire maximum MIM, which implies parametric MIM has been operational. To be specific, we contemplate the issue of the importance coefficient $\varpi$ related to $p_j$, $\varpi_{j}=F(p_j)$ satisfying
\begin{equation}
L\left ( \textbf{\emph{p} },\varpi^{*} \right ) \geq L\left ( \textbf{\emph{p} },\varpi_{j} \right )
\end{equation}

According to the analysis above, we define a objective function as $T( \textbf{\emph{p}},\varpi)=\log L( \textbf{\emph{p} },\varpi ) $.
Note that $L( \textbf{\emph{p} },\varpi )$ is the logarithm of $T( \textbf{\emph{p}},\varpi)$, we can know that they maintain the same convexity. In the light of the approach of Lagrange KKT conditions, we have the optimization problem as follow

\begin{equation}
\begin{aligned}
&\max:T( \textbf{\textsl{p}},\varpi ) = \sum\limits_{i = 1}^n {{p_i}{e^{\varpi (1 - {p_i})}}},\\
\end{aligned}
\end{equation}
\begin{numcases}{s.t.}
p_{i}-p_{i}^{(2)} < 0, & $( i=1 )$ \\
p_{i}^{(1)}-p_{i} < 0,  & $( i=1 )$ \\
\sum_{i=1}^{n}{p_{i} }=1,  & $( n=2 )$.
\end{numcases}

Resorting to the method of Lagrange with KKT conditions, we have the Lagrange function $Q$ and its partial derivative $\frac{\partial Q}{\partial p_i} $, as follow
\begin{equation}
\begin{aligned}
&Q( \textbf{\textsl{p}}, a_{j}^{(1)},a_{j}^{(2)}, \lambda) \\
&= \sum\limits_{i = 1}^n {{p_i}{e^{\varpi (1 - {p_i})}}}  + a_j^{(1)}(p_j^{(1)} - {p_j}) + a_j^{(2)}({p_j} - p_j^{(2)}) \\
&+ \lambda (\sum\limits_{i = 1}^n {{p_i}}  - 1),(n=2,j=1, a_j^{(1)}\geq 0, a_j^{(2)}\geq 0),
\end{aligned}
\end{equation}

\begin{equation}
\begin{aligned}
\frac{\partial Q}{\partial p_i}
={e^{\varpi (1 - {p_i})}}(1 - {p_i}\varpi )
- a_i^{(1)} + a_i^{(2)} + \lambda, (i=1,2).
\end{aligned}
\end{equation}
By solving $\frac{{\partial Q}}{{\partial {p_i}}}=0$ with $\sum\limits_{i = 1}^n {{p_i}}  - 1 = 0$, $a_1^{(1)}(p_1^{(1)} - {p_1})=0$, $a_1^{(2)}(p_{1} - p_{1}^{(2)})=0$, $(p_1=p, p_2=1-p)$,  it can be shown that extreme value of \textsl{Q} is obtained when $a_1^{(1)} = a_1^{(2)}=0$, $g(p, \varpi) = 0$, where,
\begin{equation}
\begin{aligned}
&g(p, \varpi)
&=(1 - p\varpi){e^{\varpi (1 - p)}}  - (1 - (1 - p)\varpi ){e^{\varpi p}}.
\end{aligned}
\end{equation}

\begin{figure}[!t]
  \centering
  \includegraphics[width=3.5in]{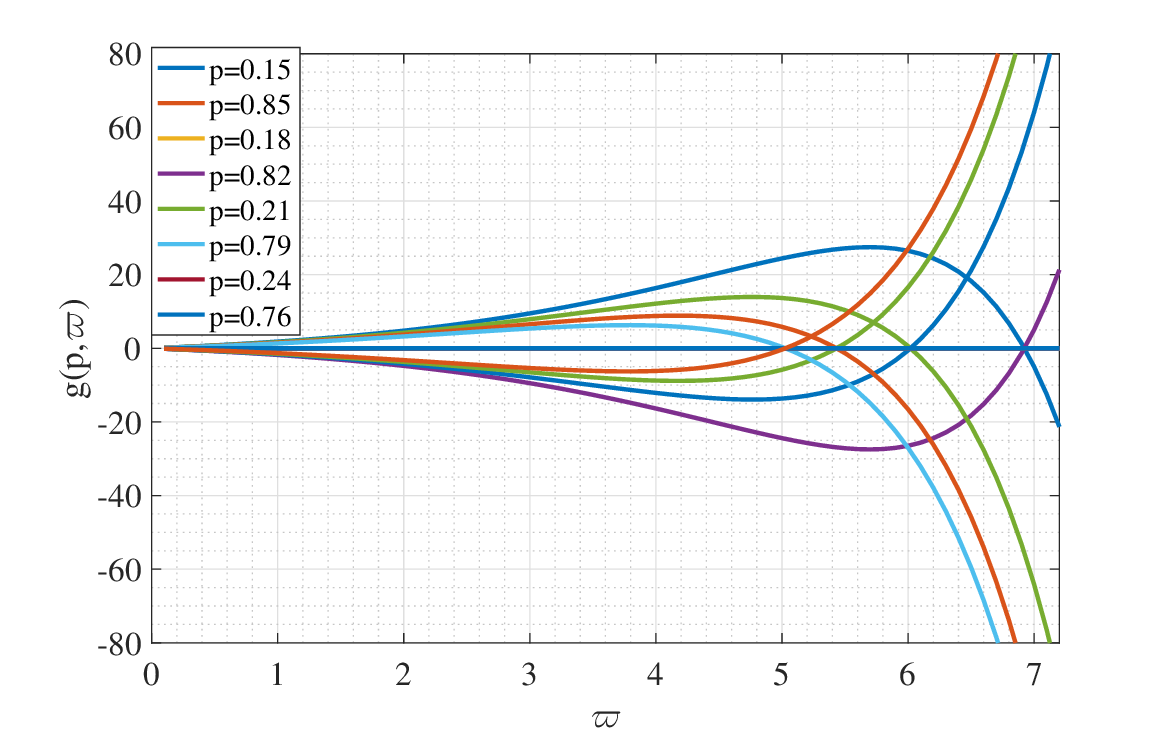}
  \caption{equation of $g(p, \varpi)=0$ with different $\varpi$ }\label{fig3}
\end{figure}

By setting $g(p, \varpi)=0$, we can derive the importance coefficient $\varpi^{*}$ contributing to the maximum MIM given the prior probability range. We then consider $\varpi^{*}$ as the selected parameter of MIM when we focus on the probability element $p$ in the binomial distribution, regarding the relationship between $\varpi^{*}$ and $p$ as the $\varpi_{j}=  F(p_j)$. Moreover, the $g(p, \varpi)$ in some certain $p$ are displayed in Fig. \ref{fig3} where we can acquire $\varpi^{*}$ the solutions of equation $g(p, \varpi)=0$ with different $p$.

In particular, $g(p, \varpi)$ can be simplified using the Taylor expansion as follow
\begin{equation}
\begin{aligned}
&g(p, \varpi)\\
&=(2p + 2) + ({p^2} - p + \frac{1}{2})\varpi  + ( - \frac{1}{2}p + \frac{1}{2}{p^2} - {p^3}){\varpi ^2}.
\end{aligned}
\end{equation}

In the light of $g(p, \varpi)=0$ and $\varpi >0$, we have
\begin{equation}
\begin{aligned}
\varpi^{*} \doteq \frac{{{p^2} - p + \frac{1}{2} + \sqrt {9{p^4} + 2{p^3} + 2{p^2} + 3p + \frac{1}{4}} }}{{2{p^3} - {p^2} + p}}.
\end{aligned}
\end{equation}

\begin{Rem}
For $\frac { \partial \left(\varpi^{*}\right) }{ \partial p}<0$, the importance coefficient $\varpi^{*}$ is monotonically decreasing with $p$. With the consideration of $p^{(1)} < p < p^{(2)}$, we can select $p=p^{(1)}$ and then determine the parameter $\varpi^{*}$ of the MIM focusing on the probability $p$ in the binary distribution with prior probability.
\end{Rem}

Furthermore, comparing $pe^{\varpi(1-p)}$ with $(1-p)e^{\varpi p}$, we have
\begin{equation}\label{}
  z(p)= pe^{\varpi(1-p)}- (1-p)e^{\varpi p},
\end{equation}
\begin{equation}\label{}
  \frac{\partial z(p)}{p}= (1-\varpi p)e^{\varpi(1-p)} + (1-(1-p)w)e^{\varpi p}.
\end{equation}

It can be derived that $\frac{\partial z(p)}{p}<0$ holds true when $ 0 \leq p \leq \frac{1}{2}$. As a result, we have
\begin{equation}\label{}
  pe^{\varpi(1-p)} \geq (1-p)e^{\varpi p}.
\end{equation}

Thence, the probability element $p$ is also the dominant when $\varpi = \varpi^{*}$.

\subsection{Bounds of Importance Coefficient with Prior Probability}
On the one hand, we can investigate the value range of the parameter $\varpi^{*}$ based on equation  $g(p, \varpi)=0$. On the other hand, we can substitute the equation $(1-p\varpi)e^{\varpi(1-p)}=C$($C$ is a constant) with more than one solution for the solution $\varpi^{*}$ of $g(p, \varpi)=0$. It is clear that the function $g(p)=(1-p\varpi)e^{\varpi(1-p)}$ is not monotonic with $p$. Therefore we look forward to explore the $\varpi$ satisfied the equation in the case of $0<p<\frac{1}{2}$. We can acquire the derivative
\begin{equation}
\begin{aligned}
\frac{{\partial g(p)}}{{\partial p}} = ({\varpi ^2}p - 2\varpi ){e^{\varpi (1 - p)}}.
\end{aligned}
\end{equation}
Considering $\varpi >0 $, the solution $\varpi$ meeting the convexity of $g(p)$ needs to make it possible to fit $\frac{\partial g(p)}{\partial p}=0$ for $0<p<\frac{1}{2}$. In consequence, we have

\begin{equation}
\begin{aligned}
\frac{2}{p_{max}} \leq \varpi \leq \frac{2}{p_{min}},
\end{aligned}
\end{equation}
where $p_{max}$=max\{$p$\}, $ p_{min}$=min\{$p$\}.

\begin{Rem}
We can provide a loose value range of the $\varpi$ depending on $0<p<\frac{1}{2}$. In view of the looser range of $p$, the parameter $\varpi$ of MIM focusing on the probability $p$ in binary distribution
belongs to \{$\varpi$ $|$ $\varpi > 4$\}.
\end{Rem}

\section{ The availability in big data}
In this section, we investigate the convergence of MIM focusing on a certain probability element under a stochastic statistical model so as to present its availability in big data. 

\subsection{Model of minority subset}
For a stochastic variable \{$X_n$\}, the sample space \{$a_1, a_2,...,a_K$\} corresponding to the probability distribution \{$p_1, p_2,..., p_K$\}. In a stochastic sequence $X_1X_2...X_M$, occurrence number of $a_k$ is $m_k$. As for $\forall \varepsilon, \delta >0 $, in the light of the weak law of large numbers \cite{Probability: theory and examples}, we have
\begin{equation}\label{law1}
\begin{aligned}
p\left\{ | \frac{ m_k }{M}- p_k | < \varepsilon \right\} > 1-\delta.
\end{aligned}
\end{equation}
It is easy to see that in general, $\frac{ m_k }{M}$, the occurrence frequency of $a_k$, is converging to the probability $p_k$ as $M$ increasing. That is to say the occurrence number of $a_k$ in the stochastic sequence can be regarded as $Mp_k$ approximately.

Meanwhile, we can also have some $m_k(k=1, 2,..., K)$ satisfying
\begin{equation}\label{law2}
\begin{aligned}
\big| \frac{ m_k }{M}- p_k \big| \geq \varepsilon,
\end{aligned}
\end{equation}
which can be considered as the condition for the minority subset $\bar{G}$. Thus, we have
\begin{equation}
\begin{aligned}
\bar{G}=\left\{ m_k \displaystyle \big| \  | \frac{ m_k }{M}- p_k | \geq \varepsilon , k=1,2,...,K \right\},
\end{aligned}
\end{equation}
\begin{equation}\left\{
\begin{aligned}
&p\{ \bar{G}\} \leq K \max\limits_{k} { p\{| \frac{ m_k }{M}- p_k | \geq \varepsilon \}} <\delta\\
&p\{G\}=1-p\{ \bar{G}\}> 1-\delta.
\end{aligned}\right.
\end{equation}

Therefore, the problem of minority subset detection can be processed as a test of binary distribution $\textbf{\emph{p}}=( p, 1-p)$ with $p(\bar{G})=p$, $p(G)=1-p$.

\subsection{The availability of parametric MIM in the model}
In order to simplify the model of the minority subset, we can focus on the binary stochastic variable, namely, $K=2$ rather than multivariable. In this case, we can regard the occurrence of $\{\bar{G_0}\}$ as a small-probability event which attracts our attention, as well as
\begin{equation}
\begin{aligned}
\bar{G_0}=\left\{ m \big| \  m \geq Mp_1+M\varepsilon, or, m \leq Mp_1-M\varepsilon \right\},
\end{aligned}
\end{equation}
where $m$ is the occurrence number of $a_1$, and the occurrence number of $a_2$ relaying on $m$.

Furthermore, we denote $n_i $ as the number of occurrence of $\{\bar{G_0}\}$ in the trial events number $N_i$, which indicates that small-probability event $\{\bar{G_0}\}$ is focused on. Thus, we have the empirical probability
\begin{equation}
\begin{aligned}
\hat{p}_{(i)}=\frac{ n_i }{ N_i }=\frac{ n_{i-1} + \Delta n_{i}}{ N_{i-1} + \Delta N_{i}},
\end{aligned}
\end{equation}
where $n_i=\sum_{j=1}^{i}\Delta n_{j}$, $N_i=\sum_{j=1}^{i}\Delta N_{j}$, $\Delta n_{j}$ and $\Delta N_{j}$ denotes the increment of $n_{j}$, $N_{j}$ respectively. We further have
\begin{equation}
\begin{aligned}
\min\{ \frac{ n_{i-1} }{ N_{i-1} }, \frac{  \Delta n_{i}}{ \Delta N_{i}} \} \leq \hat{p}_{i} \leq \max\{ \frac{ n_{i-1} }{ N_{i-1} }, \frac{  \Delta n_{i}}{ \Delta N_{i}} \},
\end{aligned}
\end{equation}
which indicates the empirical probability $\hat{p}_{i}$ will be disturbed by the increment $\frac{  \Delta n_{i}}{ \Delta N_{i}}$. However it can converge in the case of $N_{i}$ large enough based on law of large numbers.

For the MIM focusing on probability $\hat{p}_{i}$, taking the empirical probability $\hat{p}_{i}$ into
$L_i(\textbf{\textsl{p}}, \varpi_{i})$ with $\varpi_{i}=\frac{1}{\hat{p}_{i}}$, we can acquire the maximum  for  probability element $p$ based on property 2. In this case, we can have MIM with empirical probability as follow
\begin{equation}
\begin{aligned}
\hat{L}_{i} = L(\hat{p}_{i}) &= L_i\left(\hat{p}_{i}, \varpi_{i} = \frac{ 1 }{ \hat{p}_{i} } \right) \\
&=\log( \hat{p}_{i}e^{ \frac{1}{\hat{p}_{i} }-1 } +(1-\hat{p}_{i})e ).
\end{aligned}
\end{equation}
Note that the derivative satisfies $ \frac {\partial L(\hat{p}_{i}, \varpi_{i} = \frac{ 1 }{ \hat{p}_{i} } )}{\partial \hat{p}_{i}} < 0$, namely the empirical MIM $\hat{L}_{i}$ is monotonically decreasing with $\hat{p}_{i}$. Therefore we have
\begin{equation}
\min \{ \hat{L}_{i-1},\hat{L}_{\Delta i} \}
\leq \hat{L}_{i} \leq
\max \{ \hat{L}_{i-1}, \hat{L}_{\Delta i} \},
\end{equation}
where $\hat{L}_{\Delta i}= L(\frac{  \Delta n_{i}}{ \Delta N_{i}}, \varpi_{i} = \frac{  \Delta N_{i}}{ \Delta n_{i}} )$

This describes the convergence of empirical MIM $\hat{L}_{i}$ related to $\hat{p}_{i}$ when $\hat{p}_{i}$ is focused on. According to the law of large numbers, $\frac{ n_{i} }{ N_{i} }$ plays a vital role in convergence of $\hat{L}_{i}$. However, $\frac{  \Delta n_{i}}{ \Delta N_{i}}$ can disturb the convergence to cause the fluctuation.

Taking into account the property of binary distribution, we can acquire the mean $p$, and the variation $p(1-p)$. According to the central limit theorem, we derive

\begin{equation}\left\{
\begin{aligned}
&E(\hat{p}_{(i)})=p\\
&D(\hat{p}_{i})=p(1-p)/N_{i}
\end{aligned}\right.
\end{equation}
where $E(\hat{p}_{(i)})$ and $D(\hat{p}_{i})$ denote the mean $\mu$ and variation $\sigma^{2}$ of $\hat{p}_{i}$ respectively.

Note that $\hat{L}_{i}$ is a stochastic function of $\hat{p}_{i} $, we can obtain the mean $E\left( {\hat{L}_{i}} \right)$ and variation $D\left( {\hat{L}_{i}} \right)$. However, it is quite complicated to acquire the exact result, thus we can have approximate solution as follow
\begin{equation}
\begin{aligned}
&\begin{aligned}
E\left( {\hat{L}_{i}} \right)&\doteq L(\mu ) +\frac{{{L^{\prime \prime}}(\mu )}}{2}{\sigma ^2}\\
& \doteq \log \left( {p{e^{(\frac{1}{p} - 1)}} + (1 - p)e} \right)+\\
&\frac{ (\frac{2}{p}-1)e^{(\frac{2}{p}-4)}+(\frac{1}{p^3}-\frac{1}{p^2}-\frac{2}{p}+2)e^{(\frac{1}{p}-2)}-1 }
{2{{{\left( {p{e^{(\frac{1}{p} - 1)}} + (1 - p)e} \right)}^2}}}\sigma^{2}
\end{aligned}\\
&\begin{aligned}
D\left( {\hat{L}_{i}} \right) &\doteq {\left[ {{L^{\prime}}( \mu)} \right]^2}{\sigma ^2}\\
 &\doteq \left ( \frac{ (1-\frac{1}{p})e^{(\frac{1}{p}-2)}-1 }{{p{e^{(\frac{1}{p} - 2)}} + 1 - p}}\right )^{2}\sigma^{2}.\\
\end{aligned}
\end{aligned}
\end{equation}

On the other hand, in the accordance with Chebyshev inequality, we can obtain the relationship between  convergence and its probability as follow
\begin{equation}
\begin{aligned}
P\{ |\hat{L}_{i}-E( \hat{L}_{i} )| \geq \epsilon \} \leq \frac{D(\hat{L}_{i})}{\epsilon},
\end{aligned}
\end{equation}
where $\forall \epsilon>0$.

In consideration of the $E( {\hat{L}_{i}} )$ and $D( {\hat{L}_{i}} )$, we can see that when the probability element $p$ is very small, the mean $E( {\hat{L}_{i}})$ will be greatly effected by the $\frac{{{L^{\prime \prime}}(\mu )}}{2}$ except for a large enough $N_{i}$. Nevertheless, the variation $D({\hat{L}_{i}})$ can converge to $L(\mu)$ with $N_i$ increasing, especially $N_i \rightarrow \infty$, on account of the fact that $[ {{L^{\prime}}( \mu)} ]^2 $ is monotonically decreasing with $\mu$, ($0< \mu < \frac{1}{2}$).


Furthermore, it can be actually conscious that the more trial events $N_{i}$ are observed, the more exactly mean of empirical operational MIM, i.e. $E( {\hat{L}_{i}} )$, will be estimated. As for the variation, we can ascertain that the empirical MIM is actually more convergent in that condition. That is to say, it is available to take advantage of the MIM focusing on a probability element for minority subset or anomaly detection in empirical statistics.

\section{Numerical Result}
\begin{figure}[!t]
  \centering
  \includegraphics[width=3.5in]{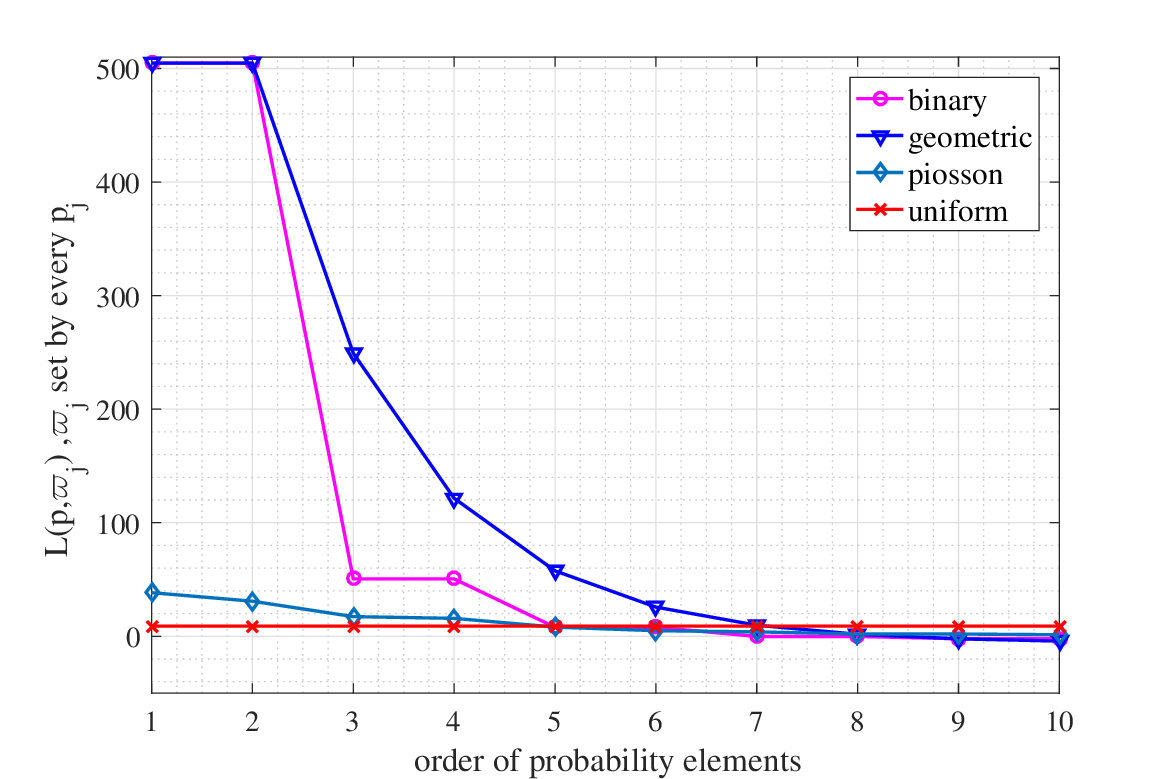}\\
  \caption{message importance measure versus $\varpi_{i}$}\label{fig1}
\end{figure}

\begin{figure}[!t]
  \centering
  \includegraphics[width=3.5in]{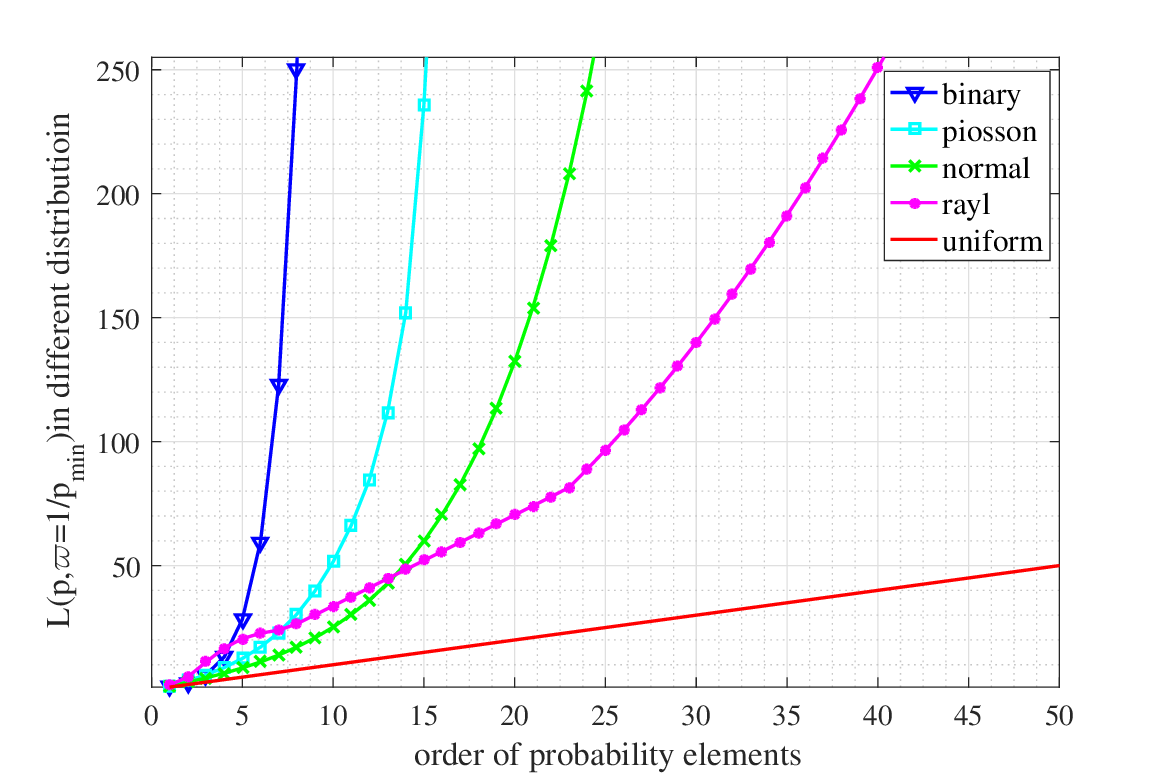}\\
  \caption{message importance measure in different distribution}\label{fig2}
\end{figure}

In the section, the properties and the availability of MIM focusing on a probability element are illustrated by numerical result.
Taking the binomial distribution, piosson distribution and geometric distribution as examples of different distributions, the variety of operational MIM with importance coefficient $\varpi_{j}$ which is set by $p_{j}$, is presented in Fig. \ref{fig1} and Fig. \ref{fig2}.

From Fig. \ref{fig1}, it is shown that the operational MIM $L_j(\textbf{\emph{p} },\varpi_{j}=\frac{1}{p_{j}})$ is decreasing with $p_{j}$ for different distributions except the uniform distribution, which would certify the property 1. Besides, it is noted that not every $\varpi_{j}$ set by $p_{j}$ in different distribution can contribute to the $L_j(\textbf{\emph{p}},\varpi_{j}=\frac{1}{p_{j}})$ larger than that in uniform distribution. However, for Fig. \ref{fig2}, it is presented that $L_0(\textbf{\emph{p} },\varpi_{0}=\frac{1}{p_{min}})$ in other distributions is bigger than that in uniform distribution, which can verify property 4.

%
%

\section{Conclusion}
In this paper, we investigated the information measures problem and discussed the parameter selection of MIM focusing on a probability element, which characterized the event uncertainty as entropy and emphasized the importance of the attractive probability element. Furthermore, the advantage of MIM focusing on a probability element is that it can choose parameter more adaptively as required for the statistical big data analysis. We have discussed several properties of a MIM with attractive probability, and investigated an practical method to select coefficient $\varpi$ with the prior probability. We also have investigated the availability of the parametric MIM with importance coefficient selected by probability for more common minority subset detection problems. In the future, we plan to design more efficient algorithms to detect anomalous events or minority subsets.


\section*{Acknowledgment}
We thank a lot for the support of the China Major State
Basic Research Development Program (973 Program) No.2012CB316100(2), and China Scholarship Council. Meanwhile, the work received the extensive comments on the final draft from other members of Wistlab of Tsinghua University.



%

\end{document}